\documentstyle[12pt,a4,epsfig]{article}
\begin{document}
\title{ \vspace*{-2cm}
Charged pions from Ni on Ni collisions \\
between $1$ and $2$ AGeV
}
\author{
D. Pelte$^6$, M. Eskef$^6$, G. Goebels$^6$, E. H\"afele$^6$,
N. Herrmann$^{4,6}$, \and M. Korolija$^{6,11}$, H. Merlitz$^{4,6}$,
S. Mohren$^6$, M. Trzaska$^6$, \and J.P. Alard$^3$, V. Amouroux$^3$,
A. Andronic$^1$, Z. Basrak$^11$, \and N. Bastid$^3$, I. Belyaev$^7$,
D. Best$^4$, J. Biegansky$^5$, A. Buta$^1$, \and R. \v{C}aplar$^{11}$,
N. Cindro$^11$, J.P. Coffin$^9$, P. Crochet$^9$, \and P. Dupieux$^3$,
M. D\v{z}elalija$^{11}$, J. Er\"o$^2$, P. Fintz$^9$, Z. Fodor$^2$, \and
A. Genoux-Lubain$^3$, A. Gobbi$^4$, G. Guillaume$^9$, \and K.D. Hildenbrand$^4$,
B. Hong$^4$, F. Jundt$^9$, J. Kecskemeti$^2$, \and M. Kirejczyk$^{4,10}$,
P. Koncz$^2$, Y. Korchagin$^7$, R. Kotte$^5$, \and C. Kuhn$^9$,
D. Lambrecht$^3$, A. Lebedev$^7$, I. Legrand$^1$, \and Y. Leifels$^4$,
V. Manko$^8$, J. M\"osner$^5$, D. Moisa$^1$, W. Neubert$^5$, \and
M. Petrovici$^1$, C. Pinkenburg$^4$, P. Pras$^3$, F. Rami$^9$, \and
V. Ramillien$^3$, W. Reisdorf$^4$, J.L. Ritman$^4$, C. Roy$^9$, \and
D. Sch\"ull$^4$, Z. Seres$^2$, B. Sikora$^{10}$, V. Simion$^1$, \and
K. Siwek-Wilczynska$^{10}$, V. Smolyankin$^7$, U. Sodan$^4$, \and
M.A. Vasiliev$^8$, P. Wagner$^9$, G.S. Wang$^4$, T. Wienold$^4$, \and
D. Wohlfarth$^5$,  A. Zhilin$^7$ \\[5mm]
The $FOPI$ Collaboration}
\maketitle
\small \noindent
$^1$ Institute for Physics and Nuclear Engineering, Bucharest, Romania \\
$^2$ Central Research Institute for Physics, Budapest, Hungary \\
$^3$ Laboratoire de Physique Corpusculaire, IN2P3/CNRS, and Universit\'e Blaise
Pascal, Clermont-Ferrand, France \\
$^4$ Gesellschaft f\"ur Schwerionenforschung, Darmstadt, Germany \\
$^5$ Forschungszentrum Rossendorf, Dresden, Germany \\
$^6$ Physikalisches Institut der Universit\"at Heidelberg, Heidelberg, Germany \\
$^7$ Institute for Theoretical and Experimental Physics, Moscow, Russia \\
$^8$ Kurchatov Institute,  Moscow, Russia \\
$^9$ Centre de Recherches Nucl\'eaires and Universit\'e Louis Pasteur,
Strasbourg, France \\
$^{10}$ Institute of Experimental Physics, University of Warsaw, Poland \\
$^{11}$ Rudjer Boskovic Institute, Zagreb, Croatia \newpage \normalsize
\begin{abstract}
Charged pions from Ni $+$ Ni reactions at $1.05, 1.45$ and $1.93$ AGeV are
measured with the $FOPI$ detector. The mean $\pi^{\pm}$ multiplicities per
mean number of participants increase with beam energy, in accordance with
earlier studies of the Ar $+$ KCl and La $+$ La systems. The pion kinetic
energy spectra have concave shape and are fitted by the superposition of
two Boltzmann distributions with different temperatures. These apparent
temperatures depend only weakly on bombarding energy. The pion angular
distributions show a forward/backward enhancement at all energies, but not
the $\Theta = 90^0$ enhancement which was observed in case of the Au $+$ Au
system. These features also determine the rapidity distributions which are
therefore in disagreement with the hypothesis of one thermal source.
The importance of the Coulomb interaction and of the pion rescattering by
spectator matter in producing these phenomena is discussed.
\end{abstract}
\section{Introduction}
This paper presents data on the $\pi^{\pm}$ emission from Ni $+$ Ni
reactions at bombarding energies $1.06, 1.45, 1.93$ AGeV. The data were
obtained with the $FOPI$ detector at $GSI$ and cover almost $4\pi$. They
are considered a supplement and extension of a previous study of the Au
$+$ Au reaction at $1.06$ AGeV which was also performed with the $FOPI$
detector \cite{fopi1}. The purpose of our new study is to systematically
investigate the dependence of the pion production on system mass and energy
in heavy-ion collisions. The present paper is therefore organized in close
similarity to ref.\cite{fopi1}.

The pion production rate in heavy-ion collisions increases rapidly with
bombarding energy. At $14$ AGeV the number of pions measured in central
collisions has reached the number of baryons contained in the participant
\cite{brau95}. At energies where the number of mesons is still smaller than
the number of baryons, calculations based on the BUU model \cite{bao91}
\cite{dani95} have shown that the number of pions reaches its final value,
which is the one measured, only at very late stages of the
reaction. At earlier times pions are partly absorbed by the nucleons with the
result that a considerable part of the baryons consists of nucleon resonances,
primarily of the $\Delta(1232)$ resonance. For the
observed properties of pions the concept of their freeze-out time and the
complex evolution of the baryon distribution are important considerations.
Pions are emitted during the total time of the reaction and thus their
phase space distribution bears the signature of the complete reaction history.
The early measurements of the pion energy spectra \cite{naga81} \cite{naga82}
showed that these are quite different from the corresponding proton spectra.
Compared to pure Boltzmann spectra with fixed temperature $T$ the former are,
at around $1$ AGeV bombarding energy, concave shaped with a mean temperature
of about $T = 70$ MeV, whereas the latter are convex shaped with a mean
temperature of about $T = 100$ MeV. This difference was first
interpreted \cite{naga82}
as due to the different source sizes at freeze-out time, later it was
realized \cite{broc84} that pions mainly originate from the decay of the
$\Delta(1232)$ resonance, and that their spectral shape is strongly influenced
by the decay kinematics. Therefore pions are very good indicators for the
excitation of such resonances. On the other hand, besides the decay channel
$\Delta \rightarrow N + n \pi$, also the competing channel
$\Delta + N \rightarrow N + N$  determines the fate of the resonances in
nuclear matter.  The time evolution of the nuclear matter distribution is,
in addition to the decay kinematics, the second decisive factor for
the pion phase space distributions.

The question of how these processes change with mass and energy of the
colliding nuclei motivated the present investigation. The comparison with
other reactions particularly include data from ref.\cite{fopi1} and from
the work
of Harris et al.\cite{harr87}. Recently corroborative data from other
experiments on pion production, performed by the $TAPS$ \cite{taps1} and
$KaoS$ \cite{kaos1} collaborations at $GSI$ have become available. Besides
these published results, the more detailed $GSI$ reports \cite{taps1}
\cite{kaos1} and conference proceedings, as e.g. ref. \cite{seng94},
are of particular interest for the present investigation.

The paper is organized as follows: The section 2 gives a short summary of
the experimental procedures. This part is shortened to the absolute
necessary since a more detailed account of these procedures was published
in ref.\cite{fopi1}. The section 3 presents the experimental results,
frequently
data from the Au $+$ Au reaction are included to illuminate the dependence
on system mass. The summary and discussion of the results on pion production
obtained so far with the $FOPI$ detector are contained in section 4.
\section{Experimental procedures}
The reaction Ni $+$ Ni was studied at nominal beam energies of
$1.06, 1.45$ and $1.93$ AGeV. The $^{58}$Ni beam was accelerated
by the $UNILAC/SIS$ accelerator combination of the $GSI$ /Darmstadt.
The duty cycle was $75\%$ with a spill length of $4$ s. The average
beam intensity corresponded to $5 \cdot 10^5$ particles per spill.
The target consisted of a $^{58}$Ni foil of $270 \mu$m  thickness. This
corresponds to an interaction probability of $0.5\%$. The average energy
loss of the $^{58}$Ni beam in this foil amounts to $0.005$ AGeV and was
neglected.

The particles produced by the Ni $+$ Ni reactions were detected by the
$FOPI$ detector which is a modular detection system with almost $4\pi$
coverage. The $FOPI$ detector is described in \cite{fopi5} \cite{fopi6}.
Of particular importance for the present investigation are the
central drift chamber $CDC$ which is mounted inside the superconducting
magnet with a solenoidal field of $0.6$ T strength, and the forward plastic
scintillation wall $PLA$. Both detector components have complete cylindrical
symmetry, the $CDC$ covers the polar angles from $32^0$ to $150^0$ in the
laboratory frame and allows pions to be distinguished from baryons. The
particle multiplicity measured by the $CDC$ is labelled $n_{\rm CDC}$,
when pions are excluded we use as quantity the measured baryonic charge
$Z_{\rm CDC}^{\rm bar}$.
The $PLA$ covers the polar range from $7^0$ to $30^0$, it allows the
separation of particles only according to their charge number $Z$. The
measured particle multiplicity in this case is labelled $n_{\rm PLA}$.
Similarly the total particle multiplicity is labelled
$n_{\rm TOT}$ with $n_{\rm TOT} = n_{\rm CDC} + n_{\rm PLA}$.
Because of the inability of the $FOPI$ detector at the time of
the experiment to identify pions for polar angles smaller than $32^0$ the
solid angle of the pion identification amounts to around $8.2$ sr. But
pions emitted into the uncovered area of the complete solid angle can be
reconstructed by employing the symmetry relation
\begin{eqnarray}
f(\Theta , \Phi) &=& f(\pi - \Theta , \pi + \Phi),
\end{eqnarray}
where $f$ may be any angle dependent observable and $\Theta,\Phi=\varphi$
refer to the $cm$ frame. This relation is valid only for symmetric reactions.
Including the remaining detection and geometrical thresholds it is estimated
that after symmetrization around $90 \%$ of the pion momentum space becomes
accessible.

In the present study we shall use the identical conventions which were used
in ref.\cite{fopi1}. All quantities which refer to the laboratory($lab$)
or target frames will be labelled by small letters. The
center of mass($cm$) or fireball frames are generally characterized
by capital letters. Exceptions are those
quantities, like the transverse momentum, which remain unchanged under the
transformation of frames. The magnitude of the transverse momentum
$p_{\rm t}$ is given in units of $p_{\rm t}^{(0)}$ where the index $^{(0)}$
indicates a normalization by the factor
$(A\cdot P_{\rm proj} /A_{\rm proj})^{-1}$
and where $A$ is the particle's mass number. Similarly the
rapidity Y in the $cm$ system is normalized by the factor
$(Y_{\rm proj})^{-1}$
and this quantity is labelled $Y^{(0)}$. In the present experiments the
normalization factors for the pion transverse momentum have the values
$9.47, 8.09, 7.02 ({\rm GeV/c})^{-1}$ for $1.06, 1.45, 1.93$ AGeV bombarding
energy, equivalently the normalization factors of the rapidities are
$1.44, 1.26, 1.12$.

The data from the drift chamber $CDC$ were analyzed by using three
different tracking algorithms as described in ref.\cite{fopi1}. The purpose
of this threefold analysis is predominantly to obtain a measure of the
systematic errors involved in the track finding and particle identification.
Because of the smaller particle multiplicities, compared to the Au $+$ Au
reaction, the systematic errors are reduced in the present case. They are
largest for the pion multiplicities where, for large impact parameters,
they become as large as $10 \%$. For other quantities, like the pion mean
kinetic energies or the widths of their rapidity distributions, they always
remain smaller than $5 \%$. In \cite{fopi1} the uncertainty of measuring
the pion transverse momenta from the curvature of the pion tracks was
determined to increase form $4 \%$ to $7 \%$ in the range $0 < p_{\rm t} <
1000$ MeV/c. From the accuracy of $\pm 2^0$ with which the forward $CDC$
limit $\vartheta = 32^0$ is identified in the particle spectra one obtains a
similar systematic uncertainty of $7 \%$ with which the pion $lab$ momenta
can be measured in forward direction.
In this paper we generally quote only one error which is the
larger one of the statistical and systematic errors. Normally the latter
determines the error.
\begin{table}
\caption{Lower multiplicity boundaries $n_{\rm TOT}(TM5)$, cross sections
$\sigma_{\rm reac}$, and deduced radii $r_0$ for the Au $+$ Au (first row)
and the Ni $+$ Ni reactions}
\begin{center}
\begin{tabular}{|c|c|c|c|}
\hline
energy   & $n_{\rm TOT}(TM5)$ & $\sigma_{\rm reac}$ &     $r_0$      \\
$(AGeV)$ &  lower limit   &     $(barns)$   &     $(fm)$     \\
\hline
1.06     &     146        & 5.50 $\pm$ 0.50 & 1.14 $\pm$ 0.05 \\
\hline
1.06     &      60        & 2.19 $\pm$ 0.24 & 1.08 $\pm$ 0.06 \\
\hline
1.45     &      64        & 2.44 $\pm$ 0.29 & 1.14 $\pm$ 0.07 \\
\hline
1.93     &      68        & 2.35 $\pm$ 0.21 & 1.12 $\pm$ 0.05 \\
\hline
\end{tabular}
\end{center}
\end{table}

\section{Experimental results}
\subsection{Charged particle multiplicities}
The multiplicity distributions of charged particles measured with
the forward scintillation detector $PLA$ and the central drift chamber $CDC$
vary only weakly for the three chosen bombarding energies. These spectra
rise for small multiplicities $n_{\rm TOT}$, their plateaus for large
multiplicities are less well developed than in case of the heavier
Au $+$ Au reaction, c.f. Fig.6 of ref.\cite{fopi1}.
It is common practice to subdivide the multiplicity distributions into 5
bins $TM1$ to $TM5$ where the upper bin $TM5$ has its lower multiplicity
boundary at half the plateau value, and where the other 4 bins have equal
multiplicity widths. These bins select event classes of certain impact
parameter ranges, with $TM1$ corresponding to the most peripheral reactions,
and $TM5$ to the most central reactions. The values which define the lower
multiplicity boundaries of the $TM5$ bins are listed in table 1. The slight
increase of this value with bombarding energy is mainly due to an increase
of the charged pion multiplicity. The 'minimum bias' situation is defined
by the condition that events are not selected according to their total
charged particle multiplicity $n_{\rm TOT}$.
\begin{figure}
\epsfxsize=12.4cm
\epsffile[0 160 600 500]{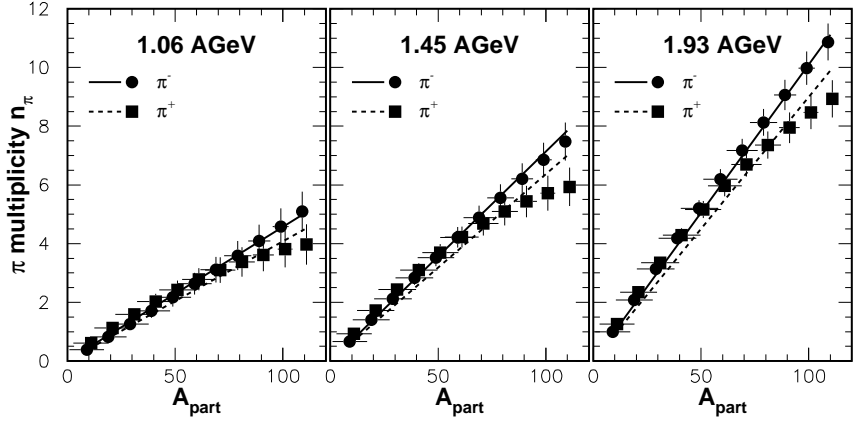}
\caption{The dependence of the $\pi^{\pm}$ multiplicities on the number of
nucleons $A_{\rm part}$ in the fireball. The full and dotted lines present
the extrapolations from the results of Harris et al.(see text)}
\end{figure}

After the
integration of the multiplicity distributions over their total range, and
after corrections because of acceptance losses for multiplicities $n_{\rm TOT}
\le 8$, the total cross sections $\sigma_{\rm reac}$ for the 3 bombarding
energies are deduced and listed in table 1. This table also includes the
corresponding information from the Au $+$ Au reaction. These cross sections
define the nuclear radii $r_0$ via $r_0 = 0.5 A^{-1/3} \sqrt{\sigma_{\rm reac}
/ \pi}$ where $A$ is the mass number of the target respectively projectile,
and where it is assumed that both nuclei have sharp surfaces. The values
of $r_0$ are listed in table 1. They are for all systems studied close to
$r_0 = 1.12 fm$ and therefore slightly smaller than the values deduced from
electron scattering \cite{hofs56}. It was argued in ref.\cite{fopi1} that
the agreement with electron scattering is improved when the smooth nuclear
surface is taken into account.

It was already pointed out that the multiplicity $n_{\rm TOT}$ is related to
the impact parameter $b$ and thus to the size of the nuclear fireball,
expressed by the number $A_{\rm part}$ of participants. To deduce the
quantitative relation between $n_{\rm TOT}$ and $A_{\rm part}$ requires to
parametrize the nuclear density distribution by a Woods - Saxon form factor
as was outlined in ref.\cite{fopi1}. Applying the identical procedures and
also using the identical global quantities
$n_{\rm TOT}, n_{\rm PLA}, Z_{\rm CDC}^{\rm bar}$,
the number of charged pions $n_{\pi^{\pm}}$ can be obtained
as function of $A_{\rm part}$. We show the dependence of $n_{\pi^{\pm}}$ on
$A_{\rm part}$ in Fig.1, where the measured pion multiplicities were used to
calculate the mean multiplicities at regular intervals of $A_{\rm part}$.
The measured pion multiplicities were in all cases multiplied with a factor
$1.1$ to correct for the remaining detection inefficiencies \cite{fopi1}.
This $10 \%$ correction of the pion multiplicities is predominantly caused
by the $\vartheta = 32^0$ cut for pion transverse momenta $p_{\rm t} < 0.6$,
c.f. Fig.3. Its size was estimated using the $IQMD$ model \cite{aich91} or a
thermal model in which pions are isotropically emitted with the measured mean
kinetic energies. Owing to these models which both do not describe the
experimental observations completely \cite{fopi1}, the accuracy of this
correction cannot be expected to be better than $50 \%$. Within this limit the
correction does not change with system mass or energy or impact parameter,
since the pion momentum distributions do not change very much, as discussed
later. The horizontal errors present the estimated accuracies
with which $A_{\rm part}$ can be obtained from $n_{\rm TOT}, n_{\rm PLA},
Z_{\rm CDC}^{\rm bar}$, the vertical errors are predominantly systematic
and present the uncertainties which were estimated by using different
trackers. In addition the Fig.1 displays the expected pion multiplicities
which one obtaines by extrapolating the Fig.2 of \cite{harr87} into the
Ni $+$ Ni system. The comparison with the expectation demonstrates
that the dependence of $n_{\pi^{\pm}}$ on $A_{\rm part}$ is not linear in
$A_{\rm part}$, but that the mean charged pion multiplicities should be close
to the results obtained by Harris et al.\cite{harr87} when extrapolated
into the Ni $+$ Ni system. In order to account for the
measured non-linearities $n_{\pi^{\pm}}$ was described by a power series of
second order in $A_{\rm part}$, i.e.
\begin{table}
\caption{Expansion coefficients $a_{\rm i}^{(2)}$ for the expansion of
$n_{\pi}(A_{\rm part})$ in powers of $A_{\rm part}$ for the Ni $+$ Ni
reactions}
\begin{center}
\begin{tabular}{|c|c|c|c|c|}
\hline
energy        &  \multicolumn{2}{|c|}{$\pi^-$} & \multicolumn{2}{|c|}{$\pi^+$}\\
(AGeV) & $a_1^{(2)}\cdot10^2$ & $a_2^{(2)}\cdot10^4$ &
           $a_1^{(2)}\cdot10^2$ & $a_2^{(2)}\cdot10^4$ \\
\hline
1.06     & $4.22\pm0.89$ & $ 0.41\pm0.71$ & $5.76\pm0.76$ & $-1.96\pm0.67$ \\
\hline
1.45     & $7.30\pm1.26$ & $-0.41\pm0.94$ & $8.73\pm1.22$ & $-3.02\pm0.95$ \\
\hline
1.93     & $11.4\pm1.53$ & $-1.41\pm1.18$ & $12.5\pm1.21$ & $-4.12\pm1.06$ \\
\hline
\end{tabular}
\end{center}
\end{table}
\begin{eqnarray}
n_{\pi}(A_{\rm part}) = a_1^{(2)} A_{\rm part} + a_2^{(2)} A_{\rm part}^2,
\end{eqnarray}
where the index $^{(2)}$ indicates the order of the expansion. The expansion
coefficients $a_{\rm i}^{(2)}$ are listed in table 2 for the three energies
used. In a way described in ref.\cite{fopi1} the average number of pions per
average number of participants can be deduced from the $a_{\rm i}^{(2)}$ via
\begin{eqnarray}
\frac{<n_{\pi}>}{<A_{\rm part}>} = a_1^{(2)} + \frac{a_2^{(2)}}{2} A_0,
\end{eqnarray}
where $A_0$ is the total number of nucleons. The results are shown in
table 3 which includes also the Au $+$ Au reaction. Notice that these values
depend on the order of the expansion. If only a
1. order expansion is chosen, as was additionally done in \cite{fopi1}, the
$\pi^-$ multiplicities in the present case were not effected, but the $\pi^+$
multiplicities would be reduced by $15 \%$. This would yield a similar increase
of the $\pi^-$ to $\pi^+$ ratios, but an $8 \%$ reduction of the total pion
multiplicities. The total pion multiplicity $<n_{\pi}>$ is calculated using
the isobar models prediction \cite{stoc86}
\begin{eqnarray}
&&n_{\pi^-} : n_{\pi^0} : n_{\pi^+}  =  \nonumber\\
&&(5N^2+NZ) : (N^2+4NZ+Z^2) : (5Z^2+NZ)
\end{eqnarray}
\begin{figure}
\epsfxsize=12.4cm
\epsffile[0 150 600 500]{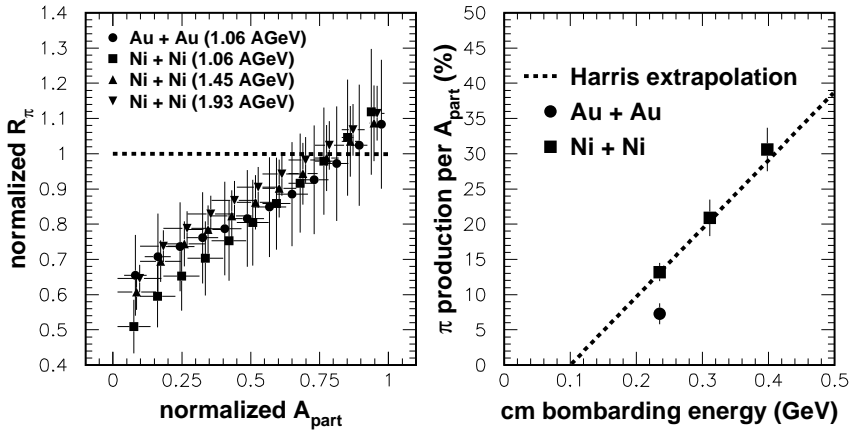}
\caption{Left: The dependence of the normalized $\pi^-$ to $\pi^+$ ratio
$R_{\pi}^{(\rm e)} / R_{\pi}^{(\rm t)}$ on the scaled number of
participants $A_{\rm part} / A_0$ for different reactions and bombarding
energies. The dotted line corresponds to $R_{\pi}^{(\rm e)} =
R_{\pi}^{(\rm t)}$. Right: The comparison between the measured mean number
of pions per participant and the extrapolation from the results of Harris
et al.(see text)}
\end{figure}
where in this case $N,Z$ are the numbers of neutrons, respectively protons
in the target or projectile. The eq.(4) holds if the pions are emitted
by a resonance with isospin $I = 3/2$, and yields for the
$n_{\pi^-}$ to $n_{\pi^+}$ ratio a value $R_{\pi}^{(\rm t)} = 1.12$. This is
in very good agreement with the experimental mean values of the ratio
$R_{\pi}^{(\rm e)} = <n_{\pi^-}>/<n_{\pi^+}>$ also shown in table 3, whereas
in case of the Au $+$ Au reaction one finds a considerable disagreement
between the model prediction $R_{\pi}^{(\rm t)} = 1.95$ and the experimental
value. If resonances with isospin $I = 1/2$ were also involved the ratios
in eq.(4) would be replaced by
\begin{eqnarray}
&&n_{\pi^-} : n_{\pi^0} : n_{\pi^+}  = 2Z : (N+Z) : 2N \hspace{5mm},
\end{eqnarray}
which is smaller than the experimentally observed value. One is, however, not
allowed to conclude that this proves a contribution from $I = 1/2$
resonances in case of the Au $+$ Au reaction, since eqs.(4,5) are valid only
for first-generation decays and do not include multiple generations of
resonances.
\begin{table}
\caption{Average number of pions $n_{\pi}$ per average number of
participants $A_{\rm part}$ and average $n_{\pi^-}$ to $n_{\pi^+}$ ratio
for Au $+$ Au (first row) and Ni $+$ Ni reactions. Only systematic errors
are shown, the statistical errors are in the order of $10\%$ of the
systematic errors}
\begin{center}
\begin{tabular}{|c|c|c|c|c|}
\hline
energy(AGeV) & $\frac{<n_{\pi^-}>}{<A_{\rm part}>}(\%)$ &
                $\frac{<n_{\pi^+}>}{<A_{\rm part}>}(\%)$ &
                  $\frac{<n_{\pi}>}{<A_{\rm part}>}(\%)$  &
                   $R_{\pi}^{(\rm e)}$  \\
\hline
1.06   & $ 3.08\pm0.37$ & $ 1.82\pm0.43$ & $ 7.26\pm0.84$ & $ 1.69\pm0.26$ \\
\hline
1.06   & $ 4.46\pm0.51$ & $ 4.62\pm0.41$ & $13.62\pm1.39$ & $ 1.19\pm0.07$ \\
\hline
1.45   & $ 7.06\pm0.80$ & $ 6.98\pm0.75$ & $21.05\pm3.29$ & $ 1.04\pm0.17$ \\
\hline
1.93   & $10.55\pm1.01$ & $10.07\pm0.79$ & $30.92\pm3.83$ & $ 1.05\pm0.13$ \\
\hline
\end{tabular}
\end{center}
\end{table}

This result appears rather puzzling because the non-linear behavior of the
pion multiplicities $n_{\pi^{\pm}}$ nevertheless yields a functional
dependence of the ratio
$R_{\pi}^{(\rm e)}$ on $A_{\rm part}$ which is very similar for all reactions
studied. In order to demonstrate this similarity we have normalized
$R_{\pi}^{(\rm e)}$ to $R_{\pi}^{(\rm t)}$ and $A_{\rm part}$ to $A_0$,
and these normalized quantities are plotted in Fig.2. After the
normalization the $\pi^-$ to $\pi^+$ ratio increases almost linearly with
the normalized $A_{\rm part}$, for small values of $A_{\rm part}$,
i.e. large impact parameters,
one finds an apparent deficit of $\pi^-$ whereas for central collisions
there appears to be a deficit of $\pi^+$. Notice that the dependence of
$R_{\pi}^{(\rm e)}$ on $A_{\rm part}$ does not disappear when $A_{\rm part}$
is determined by means of the charged particle multiplicity but employing
a different method, since a different method will not change
the values of $R_{\pi}^{(\rm e)}$ although it might change their exact
dependence on $A_{\rm part}$. Nevertheless, the condition of total charge
conservation can induce a correlation between the measured particle
multiplicity and the ratio $R_{\pi}^{(\rm e)}$ which will be discussed in more
detail in section 4.

In Fig.2 we have also plotted the dependence of the average number of all
pions per participant on the bombarding energy. As expected the Ni $+$ Ni
system closely follows the systematics obtained by
Harris et al.\cite{harr87}, whereas the pions
from the Au $+$ Au system are reduced in number by almost a factor $2$.
It was pointed out in ref.\cite{fopi1} that this reduction is confirmed in
measurements
of the $\pi^{\pm}$ production by the $KaoS$ collaboration \cite{seng94}, and
in case of the $\pi^0$ by the $TAPS$ collaboration \cite{seng94}. Both
collaborations
use detectors which have a solid angle smaller than $4\pi$ and therefore
require the extrapolation from the measured range into complete solid
angle. This extrapolation is not simple since the angular distributions
of pions are non-isotropic, c.f. section 3.2, and since the anisotropy
depends on system mass. But the
observed reduction is of a size that even a better extrapolation procedure
will not reconcile these results from different
experiments with the expectations one obtains by extrapolating from
the light-mass to heavy-mass systems.
\begin{figure}
\epsfxsize=12.4cm
\epsffile[0 150 560 450]{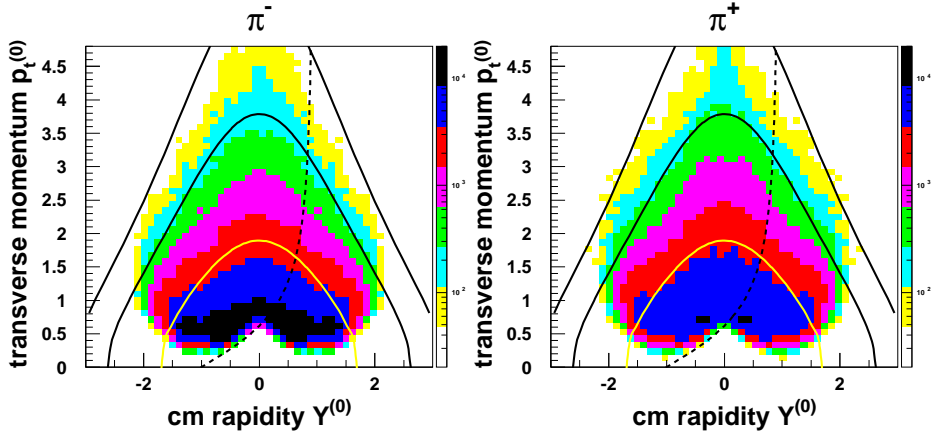}
\caption{The momentum space distribution under minimum bias condition for
$\pi^-$ and $\pi^+$ at $1.06$ AGeV bombarding energy. The full curves are
displayed for pion momenta $P = 0.2, 0.4, 0.6$ GeV/c, the dotted curves
display the $\vartheta = 32^0$ forward boundary of the $CDC$. The invariant
cross section $d^2\sigma / (p_{\rm t} dp_{\rm t} dY^{(0)}$ is displayed in
arbitrary units on a logarithmic scale}
\end{figure}
\subsection{Phase space distributions of charged pions}
In the previous investigation of the Au $+$ Au reaction the phase space
distributions of $\pi^{\pm}$ were found to be non-thermal, requiring at
least two temperatures and a non-isotropic angular distribution for their
parametrizations. Similar observations are made in the present case of the
Ni $+$ Ni reaction as will be shown in detail in the following subsections. 

The Fig.3 shows a total view onto the $\pi^-$ and $\pi^+$ momentum space
distributions under minimum bias condition in case of the Ni $+$ Ni
reaction at $1.06$ AGeV bombarding energy. Also shown are the curves of
constant momentum $P = 0.2, 0.4, 0.6$ GeV/c to illustrate the deviations
from a pure thermal behavior, and as broken curve the $\vartheta = 32^0$
geometrical limit of the $CDC$. Pions at smaller angles were not measured
but obtained by employing the reflection symmetry around $Y^{(0)} = 0$.
The Fig.3 should be compared to the
equivalent Fig.10 in ref.\cite{fopi1} to see that differences between the
pion emission from the Ni $+$ Ni and Au $+$ Au are only small. Nevertheless
such differences exist, and their detailed exposure is the subject
of the following subsections.
\subsubsection{Kinetic energies}
\begin{figure}
\epsfxsize=12.4cm
\epsffile[0 180 600 500]{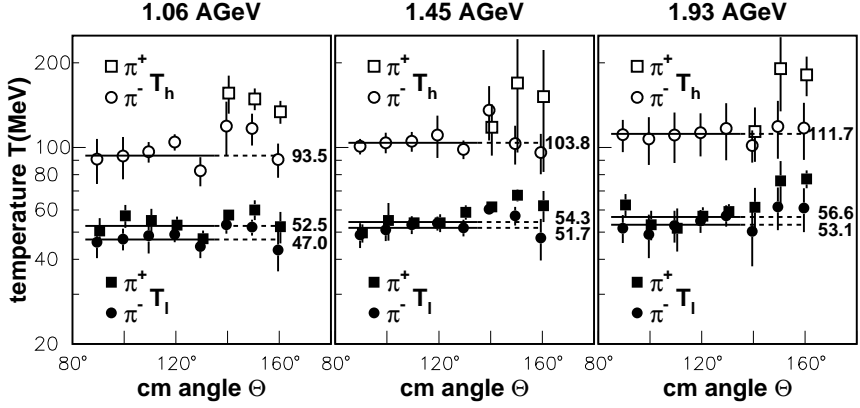}
\caption{The apparent pion temperatures $T_{\rm l,\pi}$ and $T_{\rm h,\pi}$
for different $cm$ angles $\Theta$. The lines display the the mean values
in the range $85^0 < \Theta < 135^0$}
\end{figure}
The kinetic energy distributions of $\pi^-$ and $\pi^+$ from the Ni $+$ Ni
reactions have, at all bombarding energies studied, concave shape, and
therefore two Boltzmann
distributions with different temperatures, a low temperature $T_{\rm l,\pi}$
and a high temperature $T_{\rm h,\pi}$, are needed to fit these
distributions in
the frame of a thermal picture. This is similar to the Au $+$ Au reaction
presented in ref.\cite{fopi1}, where the Fig.12 showed a representative
example of
the measured distributions at $\Theta = 130^0$. Similar to the Au $+$ Au
reaction is also that the extracted values for $T_{\rm h,\pi}$ appear to
increase for emission angles $\Theta \ge 140^0$. The angular dependence of
the low and high-temperature values is plotted in Fig.4. The values show
considerable fluctuations, the fluctuations are inherent to our procedure
which fits a
sum of two exponentials to a distribution which only slightly deviates
from a pure exponential behavior. The temperature increase at backward
angles is seen most strongly for the $\pi^+$, in order to illustrate the size
of the increase the Fig.5 compares the $\pi^-$ spectrum at $\Theta = 90^0$
to the $\pi^+$ spectrum at $\Theta = 150^0$, where the pions were measured
at $1.06$ AGeV bombarding energy.
The Fig.4 also reveals that $T_{\rm l,\pi^+}$ is systematically larger than
$T_{\rm l,\pi^-}$, whereas for $\Theta < 140^0$ the fits to the energy
distributions are consistent with the assumption that $T_{\rm h,\pi^+}$ is
equal to $T_{\rm h,\pi^-}$. The temperature values obtained from such fits
depend on the fit range. In our analysis the $\pi^-$ kinetic energy spectra
were fitted in the range from $50$ to $750$ MeV, whereas the fit range for
the $\pi^+$ kinetic energy spectra varies with angle $\Theta$: for
$\Theta = 90^0$ the range is from $50$ to $375$ MeV, the upper limit has
increased to $750$ MeV for $\Theta = 130^0$. As pointed out in
ref.\cite{fopi1} this angular variation of the fit range is necessary to
separate $\pi^+$ from protons in the analysis.
\begin{table*}
\caption{Apparent pion temperatures $T_{\rm l,\pi}$ and $T_{\rm h,\pi}$ and
mean ratio of the low-temperature to high-temperature component $R_{\rm T}$ 
for Au $+$ Au (first row) and Ni $+$ Ni reactions. The error shown
is the larger of the statistical and systematic errors for a given entry}
\begin{center}
\begin{tabular}{|c|c|c|c|c|}
\hline
energy(AGeV) & $T_{\rm l,\pi^-}$(MeV) & $T_{\rm l,\pi^+}$(MeV) &
                 $T_{\rm h,\pi}$(MeV)   &      $R_{\rm T}$         \\
\hline
1.06   & $ 42.2\pm2.7$ & $49.4\pm3.7$ &
         $ 96.4\pm5.1$ & $0.75\pm0.36$ \\
\hline
1.06   & $ 47.0\pm1.9$ & $52.5\pm3.6$ &
         $ 93.5\pm8.0$ & $1.30\pm0.81$ \\
\hline
1.45   & $ 51.7\pm1.9$ & $54.3\pm3.3$ &
         $103.8\pm4.9$ & $1.15\pm0.37$ \\
\hline
1.93   & $ 53.1\pm5.7$ & $56.6\pm4.4$ &
         $111.7\pm9.5$ & $1.06\pm0.80$ \\
\hline
\end{tabular}
\end{center}
\end{table*}
\begin{figure}
\epsfxsize=12.4cm
\epsffile[0 90 560 450]{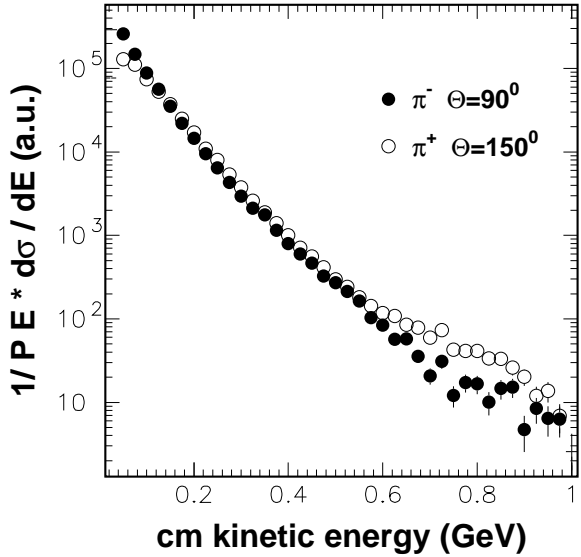}
\caption{The kinetic energy spectra of $\pi^-$ and $\pi^+$ observed at
$1.06$ AGeV bombarding energy at different $cm$ angles $\Theta$}
\end{figure}

In accordance with the analysis of the pion energy spectra performed in
ref.\cite{fopi1}, and because of the increase of the angular cross section
$d \sigma / d \Omega$ for $\Theta \ge 140^0$ (see below), we have
extracted the average pion temperatures only from pions emitted into the
angular range $40^0 < \Theta < 140^0$, the values are listed in table 4 and
shown in Fig.4. The values of $T_{\rm h,\pi}$ agree with an independent
analysis of the present data based on the transverse momentum spectra
\cite{fopi4}, but they are significantly larger $( \approx 20)$ MeV than
those obtained by the $KaoS$ collaboration \cite{kaos1}.
In the latter case the $\pi^+$ momentum spectra, measured at $\vartheta = 44^0
\pm 4^0$ in the $lab$ frame, were analyzed. Whether this difference in the
method or different fit ranges are responsible for the discrepancy is not
clear. The table 4 also contains the ratio $R_{\rm T}$, which
is the ratio between the low-temperature component $I_{\rm l,\pi}$ and the
high-temperature component $I_{\rm h,\pi}$, and it contains the equivalent
results from the Au $+$ Au reaction. The intensities $I_{\rm l,\pi}$ and
$I_{\rm h,\pi}$ were calculated by integration over the Boltzmann
distributions with temperature $T_{\rm l,\pi}$, respectively
$T_{\rm h,\pi}$. The differences between $T_{\rm l,\pi^+}$ and
$T_{\rm l,\pi^-}$, which amount in the average to
$3.9 \pm 1.5$ MeV in case of Ni $+$ Ni, but to twice this value in case
of Au $+$ Au, has its origin in the Coulomb repulsion, respectively
attraction between the pions and the positively charged baryonic matter.
Therefore the ratio $R_{\rm T}$ is influenced by the Coulomb
interaction, and to compensate this effect the values of $R_{\rm T}$ quoted
in table 4 are the mean values between $\pi^-$ and $\pi^+$. Notice that
the large errors assigned to the values of $R_{\rm T}$ are mainly systematic
and due to the fact that $R_{\rm T}$ changes more with angle for
$\Theta < 140^0$ than the temperatures.

Both pion temperatures, the low $T_{\rm l,\pi}$ temperature and the high
$T_{\rm h,\pi}$ temperature, increase slightly with beam energy. Similarly
also the relative contribution of pions with high temperatures increases,
but this component never exceeds the contribution of low-temperature
pions in the cases of the Ni $+$ Ni reactions. This is different in the
Au $+$ Au reaction where already at $1.06$ MeV bombarding energy the
high-temperature component is stronger than the low-temperature component.
\begin{figure}
\epsfxsize=12.4cm
\epsffile[80 150 540 450]{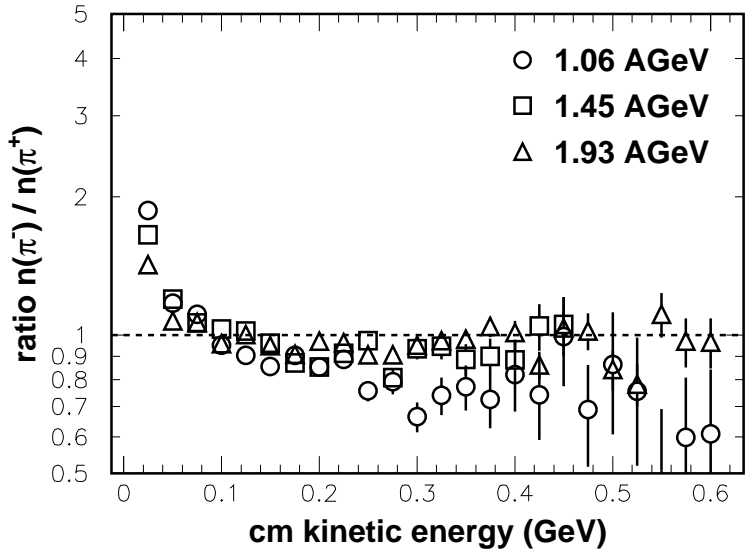}
\caption{The mean $n_{\pi^-}$ to $n_{\pi^+}$ ratio in the range $85^0 <
\Theta < 135^0$ as function of the $cm$ kinetic energy. The dotted line
corresponds to equal numbers of $\pi^-$ and $\pi^+$}
\end{figure}

The use of the term 'temperature' in discussing the kinetic energy
distributions of charged pions should not be confused with the true
thermodynamic temperature describing unordered motion. Even in the case
that pions thermalize in nuclear matter their finite mean free path adds
a zero point energy to their thermal energies because of their small mass.
The size of this quantum effect is in the order of $25$ MeV,
its general dependence on mass and localization of the pions was calculated
in ref.\cite{merl97}. In our present analysis the temperature is
solely used as a fit parameter to reproduce the concave shapes of the
energy distributions. Its small sensitivity to the bombarding energy
indicates, that the freeze-out conditions of pions do not depend strongly on
the initial conditions, on the other hand they are not completely independent
of these conditions. It it obvious that to make the differences visible, the
experimental accuracies have to be accordingly high, and the fit
conditions have to be defined in a consistent way.
\begin{figure}
\epsfxsize=12.4cm
\epsffile[0 150 600 480]{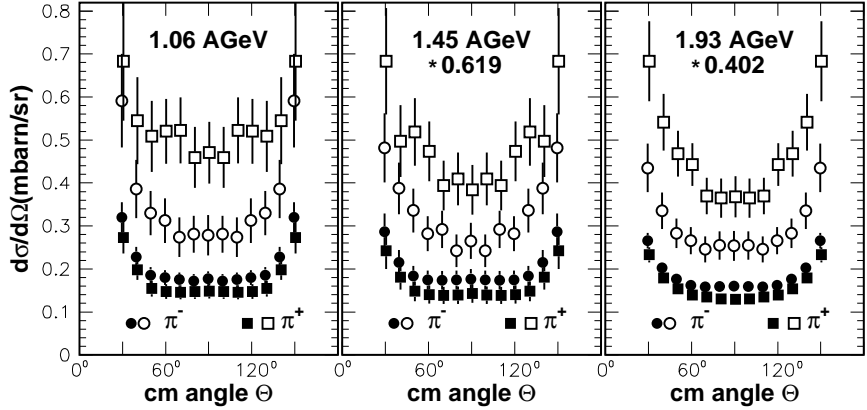}
\caption{The pion angular distributions for peripheral (open symbols) and
central (full symbols) reactions}
\end{figure}

For the sake of completeness and in accordance with the deduced pion
temperatures we display in Fig.6 the ratio of the kinetic energy spectra
of $\pi^-$ to $\pi^+$ under minimum bias conditions, where only the
$cm$ angles $90^0 < \Theta < 135^0$
are included. For the three bombarding energies studied the dependence of
this ratio on the pion kinetic energy changes only very little and is,
except for the smaller values at small kinetic energies, similar to the
Au $+$ Au case, c.f. Fig.13 in ref.\cite{fopi1}. Notice in particular that
the intersection with one occurs in all systems at kinetic energies less than
$300$ MeV. The global dominance of the $\pi^-$ over the $\pi^+$, expressed
by $R_{\pi}^{(\rm e)} > 1$ is therefore caused by the more abundant low-energy
pions, at larger energies this relation appears to be reversed. The
observation that the $\pi^-$ yield becomes smaller than the $\pi^+$ yield
for kinetic energies above $100$ MeV is linked to the reactions with large
impact parameter, c.f. Fig.1 and Fig.2. In the case of central $(TM5)$
collisions the $\pi^-$ to $\pi^+$ ratio approaches the value one for kinetic
energies above $200$ MeV. Accordingly the shape of the kinetic energy
spectra of pions, i.e. the apparent pion temperatures, do not change
significantly with impact parameter, but the yields do.
\subsubsection{Angular distributions}
The angular distributions of charged pions were found in previous
experiments to be non-isotropic with a pronounced enhancement at forward and
backward angles \cite{stoc86}. This feature is similarly observed in the
present Ni $+$ Ni reactions at all energies, c.f. Fig.7. In the Fig.7 the
measured angular distributions $d\sigma / d\Omega$ are displayed under the
selection criteria of peripheral $(TM1)$ and central $(TM5)$ collisions.
All measured pions with kinetic energies above $50$ MeV are included. Notice
that the reversal in the sizes of the $\pi^-$ and $\pi^+$ cross sections
between $TM1$ and $TM5$ is a phenomenon similarly seen in Fig.1 or Fig.2.
Since in the elementary
$N + N \rightarrow N + N + \pi$ reaction pions are also preferentially
emitted into forward and backward angles, one possible explanation for the
anisotropy found in nucleus - nucleus reactions is based on first-chance
collisions in the dilute surfaces of the colliding nuclei. Alternatively
the rescattering of pions in spectator matter yields a similar angular
anisotropy. It was argued in ref.\cite{fopi1} that this latter mechanism
could
also explain the increase of the apparent pion temperatures at forward and
backward angles. Contributions from spectators are always assumed to be
present in the phase space distributions of baryons, and a similar
anisotropy in the angular distributions of protons under minimum bias
conditions is readily interpreted as due to spectator decay. Indeed, the
forward/backward enhancement changes more strongly with impact parameter
in the Au $+$ Au reaction than in the Ni $+$ Ni reactions, and this change
is different for $\pi^-$ than for $\pi^+$. In the first reaction the anisotropy
$\sigma(150^0) / \sigma(90^0)$
increases between $TM5$ and $TM1$ by a factor $1.9 \pm 0.2$ for $\pi^-$ and
$1.4 \pm 0.2$ for $\pi^+$, whereas in the second reactions the corresponding
factors are $1.3 \pm 0.2$ and $1.0 \pm 0.2$ independent of the bombarding
energy. The differences in the surface to volume ratios for Au and Ni, and
the effects of the Coulomb attraction, respectively repulsion may be
responsible for this behaviour of the pion angular distributions. It suggests
that even in central collisions a large part of the nucleons is not stopped
in Ni $+$ Ni reactions.

The pion angular distributions reveal another peculiarity at angles around
the transverse direction $\Theta = 90^0$. Whereas pions from central
collisions of the Au $+$ Au system display a clearly discernible enhancement
of the emission into this angular range, c.f. Fig.17 of
ref.\cite{fopi1}, this enhancement is not observed
for pions from the central Ni $+$ Ni collisions at all energies studied.
Earlier measurements of the pion angular distributions from light-mass
systems gave a similar result \cite{stoc86}, i.e. a flat $d\sigma / d\Omega$
at angles
around $\Theta = 90^0$. Indeed, the transverse pion enhancement in the
Au $+$ Au central collisions constitutes the most direct evidence that the
pion phase space distributions change with system mass at bombarding
energies around $1$ AGeV. The cause for this change is not understood. It is
reminiscent of the predicted hydrodynamical flow in heavy systems when the
nuclear compression approaches values of three times normal nuclear
density \cite{stoe86}.
But it is not clear why this is observed in the pion channel when it is most
likely not present in the baryon channels \cite{fopi2}. With this respect it
should be remembered that apparent pion flow phenomena are quite often
interpreted as caused by pion absorption in the nuclear environment
\cite{bass95}\cite{bao93}.
\subsubsection{Rapidities}
\begin{table}
\caption{Parameters of the three-source fits to the pion rapidity spectra
under minimum bias condition for the Au $+$ Au (first row) and
Ni $+$ Ni reactions.
Notice that the values for the widths $\sigma$ are presented in units of
$Y^{(0)}$. The numbers in parenthesis give the systematic error in the last
2 digits. The statistical error is of the order (05) for $R_{\rm Y,\pi}$ and
(02) for $\sigma_{0,\pi}$, $\sigma_{1,\pi}$}
\begin{center}
\begin{tabular}{|c|c|c|c|c|c|c|}
\hline
energy(AGeV) &
$R_{\rm Y,\pi^{-}}$ & $\sigma_{0,\pi^{-}}$ & $\sigma_{1,\pi^{-}} $ &
$R_{\rm Y,\pi^{+}}$ & $\sigma_{0,\pi^{+}}$ & $\sigma_{1,\pi^{+}} $ \\
\hline
1.06 & 1.68(21) & 0.45(04) & 0.40(03) & 1.52(14) & 0.45(03) & 0.42(03) \\
\hline
1.06 & 1.21(05) & 0.48(01) & 0.45(01) & 1.16(07) & 0.46(01) & 0.47(02) \\
\hline
1.45 & 1.54(15) & 0.49(01) & 0.42(02) & 1.30(10) & 0.48(01) & 0.44(02) \\
\hline
1.93 & 2.68(13) & 0.51(01) & 0.38(03) & 2.43(34) & 0.53(02) & 0.39(03) \\
\hline
\end{tabular}
\end{center}
\end{table}
The forward/backward anisotropy seen in the pion angular and temperature
distributions also determines the shape of the pion rapidity distributions.
These are displayed in Fig.8, they are obtained from the complete phase
space distributions shown in Fig.3 for a special case, by a projection
onto the rapidity axis with $p_{\rm t}$ thresholds of
$p_{\rm t} > 85$ MeV/c $(1.06)$, $100$ MeV/c $(1.45)$, $115$ MeV/c $(1.93)$,
where the number in brackets specifies
the bombarding energy in units of AGeV for which these $p_{\rm t}$ cuts were
applied. The dotted curves in Fig.8 show the expected rapidity spectra, if
pions were isotropically emitted with the two temperatures $T_{\rm l,\pi}$
and $T_{\rm h,\pi}$ determined in subsection 3.2.1. It is obvious that the
rapidity distributions deviate from these expectations, the deviations
become less visible with increasing energy. In ref.\cite{fopi1} it was
assumed that
the excess intensity at rapidities around $|Y^{(0)}| = 1$ is due to
rescattering by spectator matter, and the strength of this process was
estimated by fitting the measured spectra with three Gaussian distributions
located at $Y^{(0)} = -1, 0, +1$, where the Gaussians at $Y^{(0)} = \pm1$
have to be identical for a symmetric reaction. This procedure was repeated
in case of the Ni $+$ Ni reaction. As in the Au $+$ Au case the fit
parameters, i.e. the width $\sigma_{\rm Y,\pi}$ of the Gaussians in units of
$Y^{(0)}$, and their relative intensities $R_{\rm Y,\pi} = I_{0,\pi} /
(I_{1,\pi} + I_{-1,\pi})$ do not change by more than $20 \%$ with impact
parameter. Therefore the table 5 presents the values of $\sigma_{\rm Y,\pi}$
and $R_{\rm Y,\pi}$ only under minimum bias condition, in Fig.9 the ratio
$R_{\rm Y,\pi}$ is plotted for the different impact parameters and
for the three bombarding energies of the Ni $+$ Ni reaction. In addition
we show in Fig.9 the widths $\sigma_{\rm Y,\pi}$ as functions of bombarding
energy this time in units of the unnormalized rapidity $Y$. The ratio
$R_{\rm Y,\pi}$ and the width $\sigma_{0,\pi}$ increase with bombarding
energy, the widths $\sigma_{1,\pi}, \sigma_{-1,\pi}$ stay constant within
their errors. This indicates that the number of
pions being rescattered decreases with energy, and that the separation into
3 pion sources becomes increasingly ambiguous. At energies above $2$ AGeV
the rescattered pions from the spectator sources are almost completely
masked by the non-rescattered pions from the central source. Note, however,
that the pion angular distributions still show clear evidence for the
presence of the rescattering mechanism at even the highest bombarding
energy. Compared to the Au $+$ Au reaction at $1.06$ AGeV the Ni $+$ Ni
reaction at the same energy has a stronger rescattering signal, and
this rescattering contribution to the pion rapidity spectra varies less
with impact parameter than observed in the former reaction. These
observations are the prominent indicators from the pions momentum space
distributions that the nuclear corona phenomena and the related
formation of nuclear spectators are different in the Ni $+$ Ni and
Au $+$ Au systems. And it might provide the basis on which one might
eventually find the cause for the transverse pion enhancement in central
Au $+$ Au collisions.
\begin{figure}
\epsfxsize=12.4cm
\epsffile[0 90 600 500]{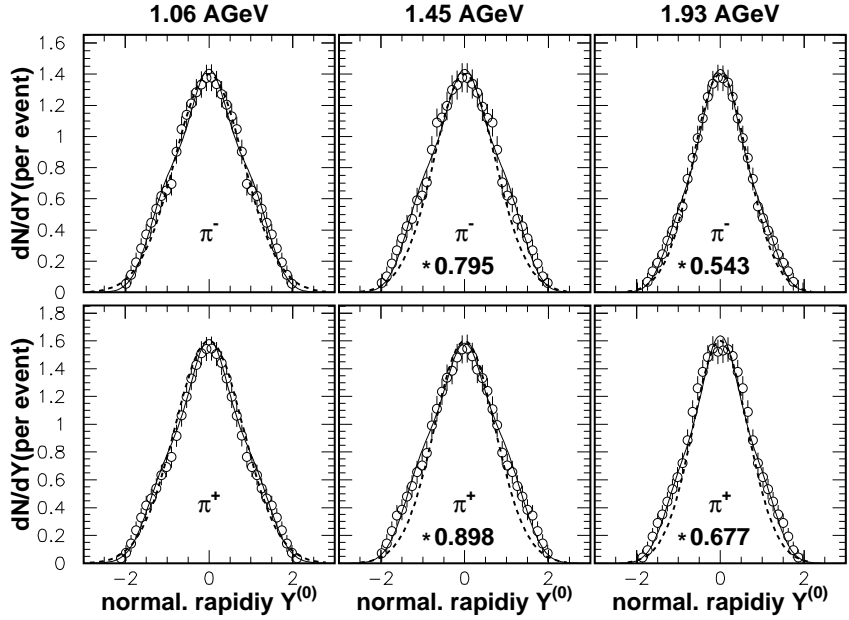}
\caption{The pion rapidity spectra at different bombarding energies. The
full curves show the fit by 3 Gaussian distributions (see text), the dotted
curves are the thermal expectations assuming two temperatures $T_{\rm l,\pi}$
and $T_{\rm h,\pi}$}
\end{figure}

\section{Summary and discussion}
The results of this investigation should be discussed in relation to the
study of pion production in the Au $+$ Au system which contains almost four
times as many nucleons. The total number of pions produced at
a given energy is therefore larger in the heavier system, but when
normalized to the
number of participants the Ni $+$ Ni system produces almost two times more
pions at $1.06$ AGeV than the Au $+$ Au system. Furthermore the Ni
result is in agreement with equivalent results from the Ar $+$ KCl and the
La $+$ La systems studied by Harris et al.\cite{harr87}. The surface to
volume ratio varies from the Ni to the La system $(1.34)$ more than from
the La to Au system $(1.12)$, and thus it is difficult to attribute the
reduction of the $\pi^{\pm}$ yield in the Au $+$ Au case to an enhanced
pion absorption in the larger fireball volume. To decide whether or not
absorption
is responsible for the reduced $\pi^{\pm}$ yields, the independence of
the normalized number of pions for system masses below $A_0 = 280$ should
be confirmed, e.g. for the $A_0 = 200$ system. Considering the fact that
the pion angular distributions depend on the system mass, it is not
sufficient to do this test at midrapidity but it should be conducted over
the complete solid angle.
\begin{figure}[t]
\epsfxsize=12.4cm
\epsffile[0 90 600 500]{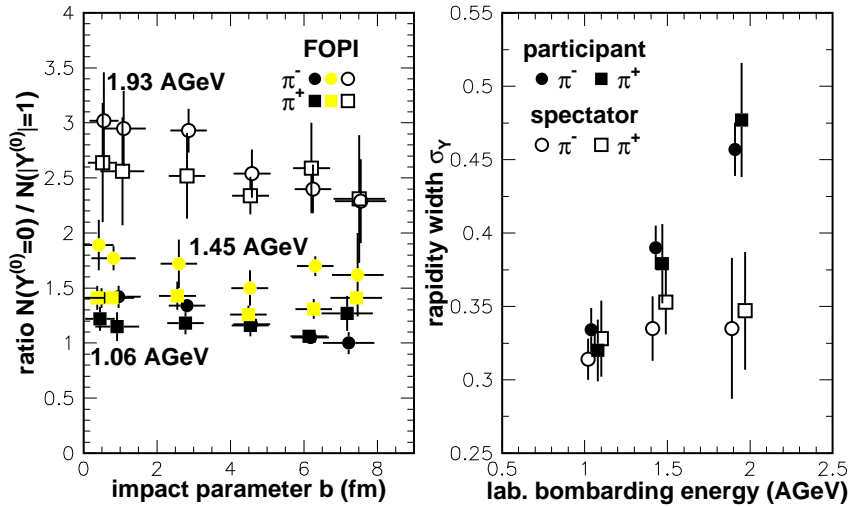}
\caption{Left: The impact parameter dependence of the intensity ratios
$R_{\rm Y}$ between pions from the fireball and pions from the spectators.
Right: The energy dependence of the widths $\sigma_{\rm Y}$ of the pion
rapidity distributions from the fireball and the spectators under minimum
bias conditions}
\end{figure}

The alternative explanation for the reduced production of pions in the
Au $+$ Au system might be an enhanced flow of energy into collective motion
which was recently observed in central collisions, although at lower
energies \cite{fopi2}. This will decrease the energy available to excite
baryon resonances. At energies around $1$ AGeV pions are considered to
mainly originate from the decay of heavy baryon resonances, where
the $\Delta(1232)$ resonance dominates the pion production. Nevertheless,
theoretical studies have shown that resonances up to a mass of
$1.9$ GeV/c$^2$ have to be included to explain the pion spectra
measured in Au $+$ Au reactions \cite{teis96}. The excitation of such
resonances is seen in the concave shape of the $\pi$ kinetic energy spectra. 
The apparent higher temperature $T_{\rm h,\pi}$ is a signal for the presence
of resonances above the lowest lying $\Delta(1232)$ resonance, as will be
discussed in more detail in a forthcoming paper \cite{fopi3}. With this
respect it is interesting to notice that $T_{\rm h,\pi}$ is larger in the
Au $+$ Au reaction than in the Ni $+$ Ni reaction at the same energy, and
that also the contribution of the high-temperature component is larger in
the Au than in all the Ni cases studied. This is a puzzling result,
since an enhanced excitation of the higher baryon resonances should lead
to an increase in the number of pions, because the $2\pi$ decay channel
becomes more important. The conflict can be resolved in a simple way,
if baryon resonances are only excited via $N + N$ collisions in the early
compression phase of the collision. In the later expansion phase their
decay by pion emission will produce second and subsequent generations of
resonances, not only in the fireball but also in the cold spectator matter.
It is reasonable to assume that the number of resonance
generations is larger in the heavier system, and that the competing
$N + B \rightarrow N + N$ channel causes a loss of pions.
In this case the total pion reduction would be proportional to a power
law with the number of generations in the exponent.

Coulomb effects are present in the pion spectra, most indicative is the
difference between the values of the low temperature $T_{\rm l,\pi^-}$ and
$T_{\rm l,\pi^+}$. This difference is only half as large in the Ni $+$ Ni
reaction than in the Au $+$ Au reaction, and appears to be independent of
bombarding energy. On the other hand the $\pi^-$ to $\pi^+$ ratio also
depends on the size of the fireball, expressed by the number of
participants $A_{\rm part}$. And this dependence, when properly normalized,
is found to be independent of the system mass or energy. It has a universal
dependence with too few $\pi^-$ at peripheral reactions, and too few $\pi^+$
at central reactions. The latter might be due to the condition that the total
charge is conserved. This condition and the requirement that for minimum
impact parameter the measured particle multiplicity attains its maximum value
selects events with large numbers of $\pi^-$, i.e. with large
$R_{\pi}^{(\rm e)}$ values. It is easy to see that the two conditions
\begin{eqnarray}
Z^{\rm bar} + (1 + R_{\pi}^{(\rm e)}) n_{\pi^+}  =  max \nonumber\\
Z^{\rm bar} + (1 - R_{\pi}^{(\rm e)}) n_{\pi^+}  =  const \nonumber
\end{eqnarray}
are the better fulfilled the larger $R_{\pi}^{(\rm e)}$.  Of course,
$R_{\pi}^{(\rm e)}$ in these equations cannot grow unlimited, it is bound by
isospin and baryon conservation. But it is conceivable that it becomes largest
when the number of measured particles is largest.
In the case of peripheral reactions the
pion rescattering process in spectator matter is probably non negligible
in determining the $\pi^-$ to $\pi^+$ ratio. The impact parameter
dependence of the pion angular distributions is suggestive: In both
systems, i.e. Au $+$ Au and Ni $+$ Ni, the forward/backward enhancement in
peripheral reactions is stronger
for $\pi^-$ than for $\pi^+$. Interpreted as due to rescattering,
the rescattering process in spectator matter would then be stronger for
$\pi^-$ than $\pi^+$, and the Coulomb attraction between the
$\pi^-$ and the positively charged spectators offers a possible
explanation for this preference. The preferred
absorption of $\pi^-$ in spectator matter, without reemission because of
the $N + B \rightarrow N + N$ channel, would also mean that pion absorption
is an important process to heat the spectators. In addition the preferred 
absorption of negative pions should cause the decay products of the
spectator to become more neutron-rich. The effects the Coulomb potential has
on the ratio $R_{\pi}^{(\rm e)}$ was recently studied in a theoretical paper
by Teis et al. \cite{teis97}. Whereas its dependence on the pion kinetic
energies, c.f. Fig.6, is quite well reproduced, the dependence on the number
of participants $A_{\rm part}$ deduced from the measured particle
multiplicities is not studied in similar detail.

With respect to the pion 'temperatures' $T_{\rm l,\pi}$ and $T_{\rm h,\pi}$
listed in table 4, we want to reiterate that these should be only
considered as fit parameters to the concave shapes of the kinetic energy
spectra. Their values depend on the chosen fit range. If the fit range
is increased it is very probable that, to obtain a good fit, also the
multitude of temperatures has to be increased. This is the reason why the
upper energy boundary in case of the $\pi^-$ spectra was limited to $750$
MeV. The other reason is that the exact shape of the kinetic energy
spectra at large kinetic energies, when the cross section has dropped by
several orders of magnitude, depends on an accurate estimate of the
background and on the way pions are selected from the data. The
identification of pions with large energies, i.e. large momenta, requires
to determine these momenta with high precision by the tracking algorithm.
Above $750$ MeV the standard deviation of the $lab$ momentum in
forward direction becomes larger than $7 \%$ \cite{fopi1} and this is
considered
inadequate to identify such rare pions with the required precision.
Nevertheless, the systematic variations of $T_{\rm l,\pi}$ and
$T_{\rm h,\pi}$ with system mass and energy allows one to make important
conclusions. If the observed pions are assumed to be exclusively remnants
from the decay of baryon resonances, then the increase of
$T_{\rm l,\pi}$ and $T_{\rm h,\pi}$
with increasing energy and mass is related to the change of the number
of resonances involved and on their excitation profile at pion freeze-out
times. In principle the pion energy spectra can be used to derive
information about these quantities. This line of analysis shall be followed
in a forthcoming paper \cite{fopi3}. On the other hand if the high-temperature
component is interpreted as due to pions thermalized in the nuclear
medium \cite{e814}, the values of $T_{\rm h,\pi}$ allow one to deduce the
fireball temperatures at pion freeze-out times. This line of analysis is
followed in ref.\cite{fopi4}. However, one should not forget the problems
which are caused by the quantum nature of pions and which were mentioned
in subsection 3.2.1. \\[5mm]
\noindent {\bf Acknowledgement}
This work was supported in part by the
Bundesministerium f\"ur Forschung und Technologie under contract
06 HD 525 I(3) and by the \linebreak Gesellschaft f\"ur Schwerionenforschung
under contract HD Pel K.

\end{document}